\documentclass{ws-ijmpa}
\usepackage{graphicx,psfrag,amsmath,amssymb,array}
\usepackage{latexsym,marvosym,color,setspace,bbm}

\makeatletter

\begin{document}

\markboth{Manfried Faber and Roman H\"ollwieser}
{Distribution of Magnetic Monopoles within cubes in Compact QED}

%
\catchline{}{}{}{}{}
%

\title{Distribution of Magnetic Monopoles within cubes in Compact QED}

\author{Roman H\"ollwieser\footnote{hroman@kph.tuwien.ac.at}\; and Manfried Faber\footnote{faber@kph.tuwien.ac.at}}

\address{Atomic Institute, Vienna University of Technology,\\ Wiedner Hauptstr.\ 8-10, A-1040 Vienna, Austria}

\maketitle

\begin{abstract} 
Earlier investigations~\cite{Bozkaya:2004tb} showed local minima in the monopole-antimonopole potential in $U(1)$ gauge theory on the lattice. In this paper we localize monopoles of Monte-Carlo configurations. A statistical analysis of localization measurements gives us the probability density which we compare with the potential found in Ref.~\cite{Bozkaya:2004tb}. We find the monopoles mainly located either in the center of three-dimensional cubes or on the interface between two cubes. This agrees with the position of minima and maxima of the monopole-antimonopole potential. 
\end{abstract}

\ccode{PACS numbers: 11.15.Ha, 14.80.Hv}

\section{Introduction}

In compact quantum electrodynamics (QED) there is a phase-transition on the space-time lattice, which separates a strongly coupled phase with confinement and a weakly coupled Coulomb-phase. The phase-transition is associated with the occurrence of topological excitations which can be identified as magnetic monopoles. The monopole condensate causes the creation of electric flux tubes via a dual Meissner effect in the abelian case, leading to a linear rising potential and therefore confinement. Although the confinement mechanism in QED seems to be rather different from the one in QCD, the investigation of the former gives some inference to the confinement in QCD, which is far more complex to investigate due to its non-abelian structure.

In Ref.~\cite{Bozkaya:2004tb} the influence of the granularity of the lattice on the potential between monopole and antimonopole was investigated. It showed periodic deviations from the $1/r$-behavior of the monopole-antimonopole potential leading to local extrema. We suppose that these properties of the potential may influence the localization of magnetic monopoles and therefore the order of the phase transition in compact QED. The order of this phase transition is of vital importance for the continuum limit, which can be obtained only in the case of a second order transition.

\section{Dirac Monopoles}

Magnetic Monopoles were introduced by P.A.M. Dirac~\cite{Dirac:1931kp}, in order to symmetrize Maxwell's equations. Using the common vector potential $A_\mu(x)$, so called Dirac strings appear which connect magnetic sources and drains. Dirac showed, that the field of these strings is invisible if the magnetic flux along the string is some integer multiple of $2\pi/e$ (in natural units) and the wave-function for a charged particle that interacts with the monopole vanishes along the string. Magnetic monopoles are quantized singularities of the gauge field and the elementary monopole charge $g$ obeys $eg=2\pi$. The existence of magnetic monopoles implies the quantization of the elementary electric charge $e$. There are two kinds of divergences, the ``true'' physical divergences of monopoles and the unphysical ``gauge'' divergences of the Dirac strings. 

For the Monte-Carlo calculations we use the Wilson action~\cite{Wilson:1974sk} of compact QED on an Euclidean 4D-lattice, given by
\begin{equation}\label{wilson}
S_W=\beta\sum_{x,\mu<\nu}(1-\cos\theta_{\mu\nu}(x)),\quad\beta=1/e^2.
\end{equation}
Because of the $2\pi$-periodicity of the links $\theta_\mu(x)\in(-\pi,\pi]$ the plaquette angle
\begin{equation}\label{plaquetteangle}
\theta_\Box=\theta_{\mu\nu}(x) = \theta_\mu(x)+\theta_\nu(x+\hat{\mu})-\theta_\mu(x+\hat{\nu})-\theta_\nu(x)\in(-4\pi,4\pi]
\end{equation}
has no direct physical meaning. 

A common choice for the definition of the field-strength is
\begin{equation}
ea^2F_\Box=\bar\theta_\Box
\end{equation}
where $a$ is the lattice constant and the physical angle $\bar\theta_\Box\in(-\pi,\pi]$ is obtained by splitting off the number of Dirac strings $n_\Box(x)\in\{-2,-1,0,1,2\}$ penetrating a plaquette 
\begin{equation}
\theta_\Box(x)=\bar\theta_\Box(x)+2\pi n_\Box(x).
\end{equation}
De Grand and Toussaint~\cite{DeGrand:1980eq} proposed to identify monopoles (antimonopoles) by counting Dirac strings starting (ending) in cubes $C$. The monopole charge $m$ in units of $g$ is then given by 
\begin{equation}
m(x) = \sum_{\Box\in\partial C} n_\Box(x),
\label{eq:degrandtoussaint}\end{equation}
where $\Box$ runs over the plaquettes enclosing the cube $C$ at point $x$. Identifying monopoles this way allows to count the number of monopoles in each cube only. For the exact localization of monopoles within cubes we have to choose another method.

Describing the flux through a plaquette by~\cite{Lang:1986ry,Skala:1996ar}
\begin{equation}
ea^2F_\Box=\sin\theta_\Box
\end{equation}
gives a continuous definition of the field-strength which takes into account the $2\pi$-periodicity. This definition is achieved by a variation of the Wilson action and is therefore in agreement with the Gau{\ss} law on the lattice~\cite{Zach:1995ni}. The magnetic charge density $\rho_m(\vec{r})$ is given by 
\begin{equation}
\mbox{div} \vec{B}(\vec{r}) = \rho_m(\vec{r}).
\end{equation}
The magnetic charge in a cube $C$ at position $x$ therefore reads
\begin{equation}
\begin{aligned}Q_m(x)&=\int_C \rho_m(\vec{r})d^3r =\int_C \mbox{div}\vec{B}(\vec{r})d^3r = \oint_{\partial C}\vec{B}(\vec{r})d^2\vec{f}\\ &= a^2\sum_{\Box\in\partial C} F_\Box = \frac{1}{e}\sum_{\Box\in\partial C}\sin\theta_\Box = \frac{g}{2\pi}\sum_{\Box\in\partial C}\sin\theta_\Box.
\end{aligned}
\label{eq:sinusflux}\end{equation}
The monopoles identified this way are still point-like but the discretization of the magnetic flux vanishes. Nevertheless, both monopole definitions give qualitatively similar results~\cite{Hoellwieser:2006}.

\section{Localization of Monopoles in Cubes}\label{pos}

\subsection{Theoretical Aspects}\label{posrech}

According to De Grand and Toussaint (\ref{eq:degrandtoussaint}) or using the sinus-flux definition (\ref{eq:sinusflux}) we can locate monopoles in certain cubes. To specify this location we use the following idea:

For a monopole in the center of a cube one plaquette occupies a solid angle $\Omega_\Box=4\pi/6$. The closer the monopole moves to the center of the plaquette, the more grows the associated solid angle, reaching $\Omega_\Box=2\pi$ for a monopole in the center of the plaquette. Assuming that the field of a monopole in the immediate surrounding of the center is spherical symmetric, the plaquette angle $\theta_\Box$ is proportional to $\Omega_\Box$, $\theta_\Box=\Omega_\Box/2$. 

\begin{figure}[h]   
\centering   
\psfrag{3.14159}[0][cr][1][0]{\small $\pi$}   
\psfrag{ 0}{\small $0$}   
\psfrag{ 0.5}{\small $0.5$}   
\psfrag{ 1}{\small $1$}   
\psfrag{r}{\small \hspace{-2mm}d/a}   
\psfrag{Fluss}[0][Br][1][0]{\small \vspace{15mm}flux $\theta_\Box(d)$}   
\psfrag{ 0}[0][Bl][1][0]{$0$}   
\psfrag{ 3.14159}[0][Bc][1][0]{\small $\pi$}   
\psfrag{-3.14159}[0][Bc][1][0]{\small $-\pi$}   
\psfrag{-1.0472}{}   
\psfrag{-2.0944}{}   
\psfrag{ 1.0472}{}   
\psfrag{ 2.0944}{}   
\psfrag{rechts}[20][Bc][1][0]{\small flux $\theta_\Box^+$}   
\psfrag{links}[-15][Br][1][0]{\small flux $\theta_\Box^-$}   
a)\includegraphics[scale=0.6]{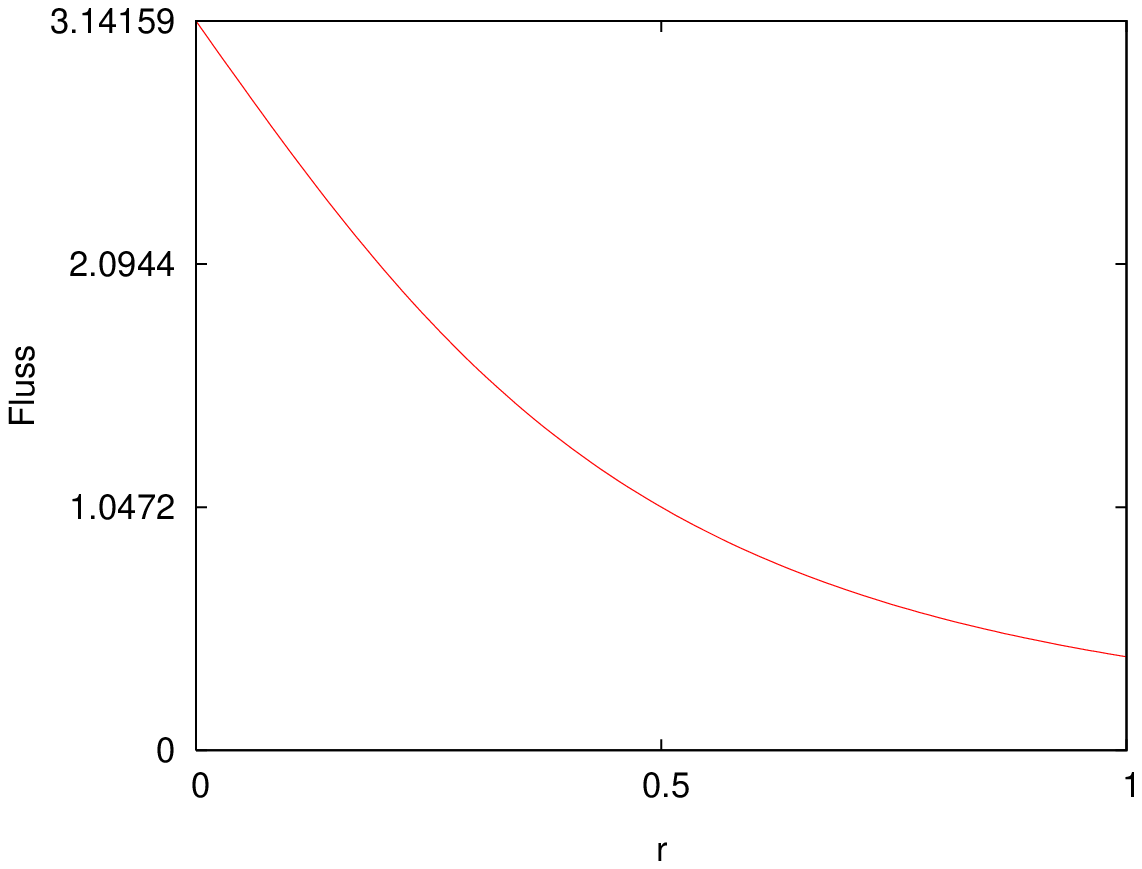}$\;$
b)\includegraphics[scale=0.57]{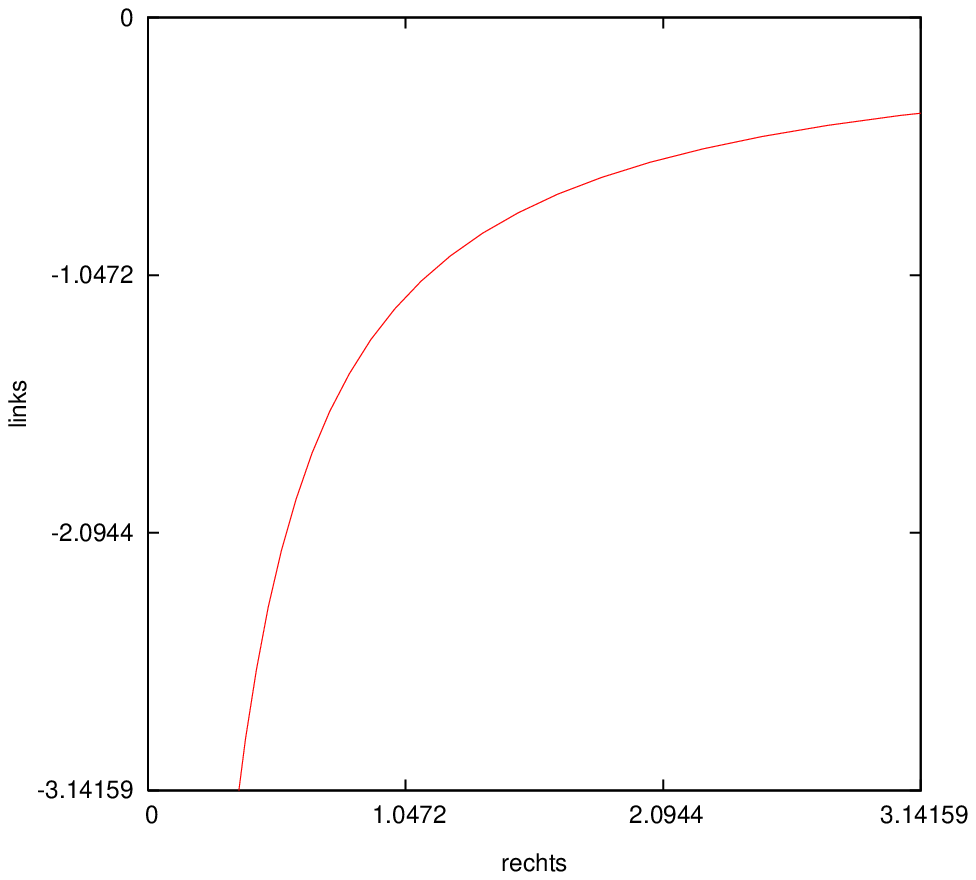}   
\caption{Flux $\theta_\Box(d)$ a) through a plaquette of a classical monopole in distance $d$ (on the plaquettes surface normal), b) through opposite plaquettes $\theta_\Box^\pm$, produced by a monopole crossing a cube.}\label{fig:flux} 
\end{figure}

This allows to determine the distance $d$ between the monopole and the plaquette from the plaquette angle $\theta_\Box$. The four unit vectors from the center of the monopole to the corners of the plaquette define a spherical quadrangle and the corresponding solid angle $\Omega_\Box(d)$ depending on the distance $d$. The flux amounts therefore to
\begin{equation}\label{fluxeq}   
\theta_\Box(d)=\frac{\Omega_\Box(d)}{2}=\pi-2\arccos\frac{1}{4(d/a)^2+1}.
\end{equation}\label{eq:flux}
The flux $\theta_\Box(d)$ through a plaquette in distance $d$ of the monopole is shown in Fig.~\ref{fig:flux}a. Negative plaquette values indicate either that the monopole is located towards smaller coordinate values or the presence of an antimonopole, which leads to negative flux differences between opposite plaquettes. Comparing the flux through opposite plaquettes $\theta_\Box^\pm$ we determine the relative distances to the plaquettes. From the three pairs of opposite plaquettes (left-right, front-back, up-down) in the cube we get the position of the monopole in the cube. A monopole located directly on a plaquette ($d=0$) produces the maximum flux, $\theta_\Box(0)=\theta_\Box^-=\pi$. According to Eq.~(\ref{fluxeq}) the flux for the opposite plaquette is still about $\theta_\Box(a)=\theta_\Box^+=\pi-2\arccos(1/5)=0.4$. For a  monopole in the center of a cube, we should measure fluxes $\theta_\Box^\pm=(\pi/3,-\pi/3)$ through opposite plaquettes. Displaying the flux through opposite plaquettes in a diagram, we expect the path shown in Fig.~\ref{fig:flux}b for a monopole crossing a cube.

\subsection{Measurements}\label{posmess}
\begin{figure}[h]   
\centering  
\psfrag{ 12.5664}[0][Bc][1][0]{$4\pi$}   
\psfrag{ 6.28319}[0][Bc][1][0]{$2\pi$}   
\psfrag{-12.5664}[0][Bc][1][0]{$-4\pi$}   
\psfrag{-6.28319}[0][Bc][1][0]{$-2\pi$}   
\psfrag{ 0}[0][Bl][1][0]{$0$}   
\begin{tabular}{cccc}     
a)\vspace{3mm}\parbox[b]{1mm}{$\quad\theta_\Box^-$\\ \\ \\ \\ \\}&\includegraphics[scale=0.52]{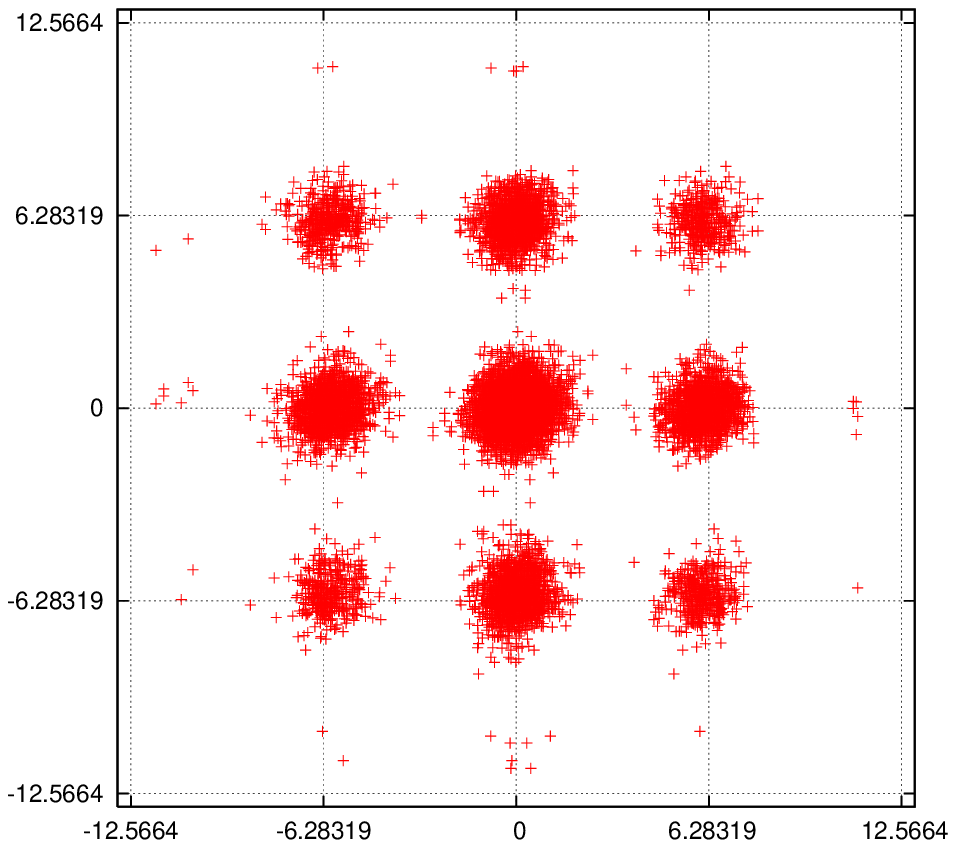}&b)\parbox[b]{1mm}{$\quad\theta_\Box^-$\\ \\ \\ \\ \\}&\includegraphics[scale=0.52]{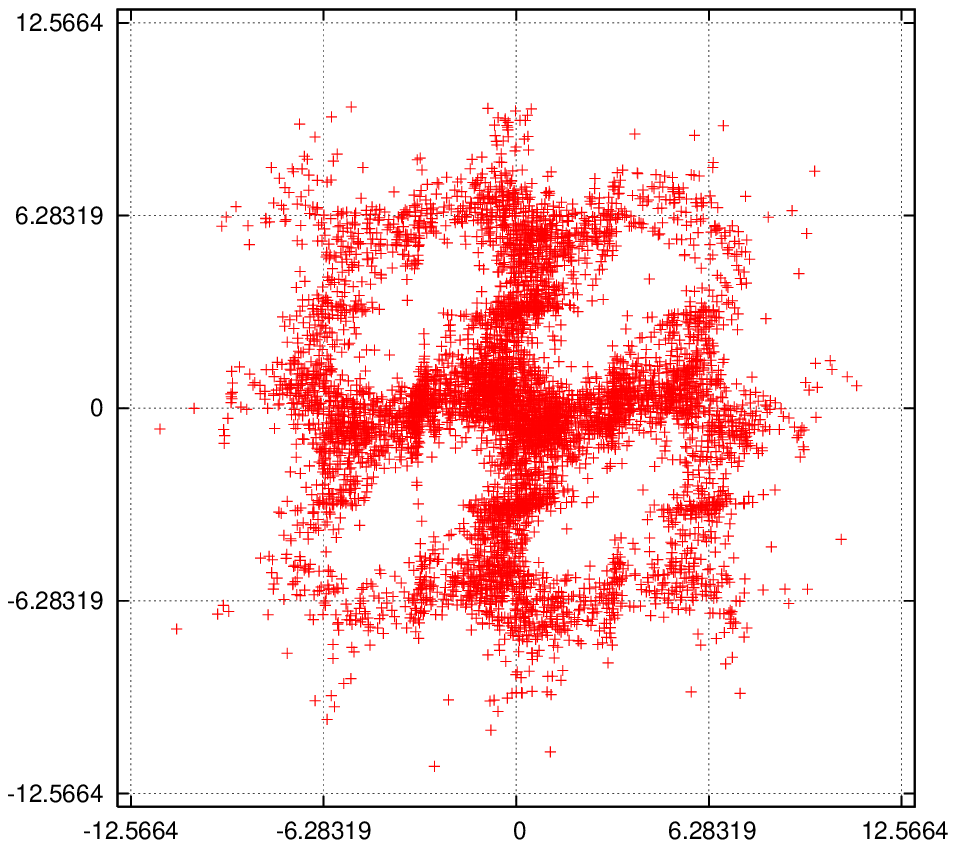}\\
&$\qquad\;\theta_\Box^+$&&$\qquad\;\theta_\Box^+$
\end{tabular}   
\caption{Distributions of opposite plaquette angles $\theta_\Box^\pm$ for a) cubes without Dirac monopoles and b) $\pm 1$-Dirac monopoles. Besides the different structures notice a $2\pi$ periodicity, indicating the presence of Dirac strings.}\label{fig_gmpsxy}
\end{figure}

To get an information about the position of monopoles inside a cube we measure opposite plaquette angles $\theta_\Box^\pm$ on $20$ Monte Carlo configurations for $\beta=0.5$ and $200$ for $\beta=1.4$ on $20^4$-lattices. In Fig.~\ref{fig_gmpsxy} we display the distribution of plaquette pairs ($\theta_\Box^+,\theta_\Box^-$) for cubes without a Dirac monopole to the left and ``$\pm1$-Dirac monopoles'' to the right at $\beta=1.4$.

\begin{figure}[h!]   
\psfrag{pi}[B][l][1][0]{$\pi$}   
\psfrag{-pi}[B][l][1][0]{$-\pi$}   
\centering   
\begin{tabular}{cccc}     
\vspace{-3mm}
a)\parbox[b]{1mm}{$\;\theta_\Box^-$\\ \\ \\ \\ \\ \\}&\includegraphics[scale=0.25]{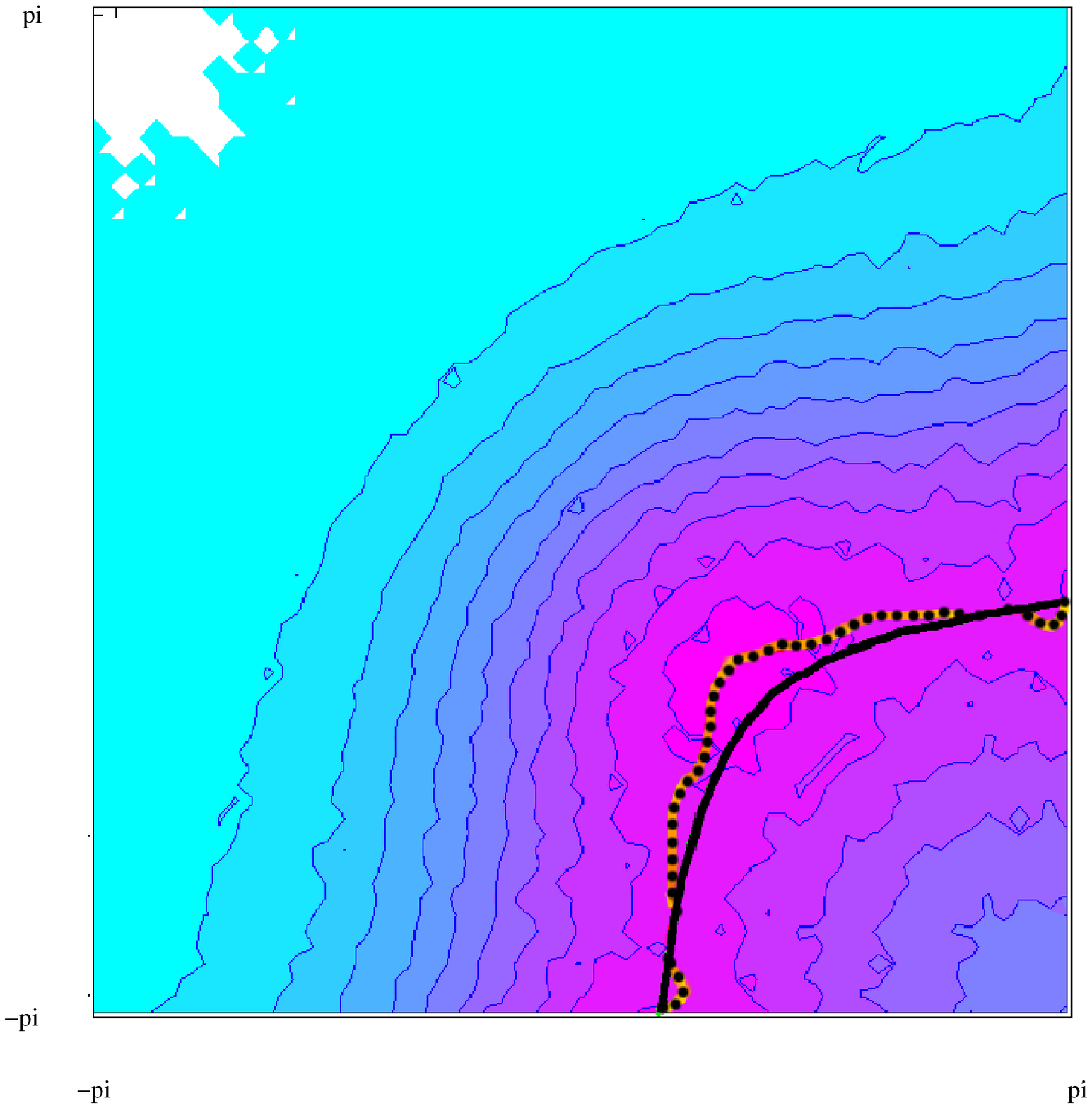} &b)\parbox[b]{2mm}{$\;\;\theta_\Box^-$\\ \\ \\ \\ \\ \\} & \includegraphics[scale=0.25]{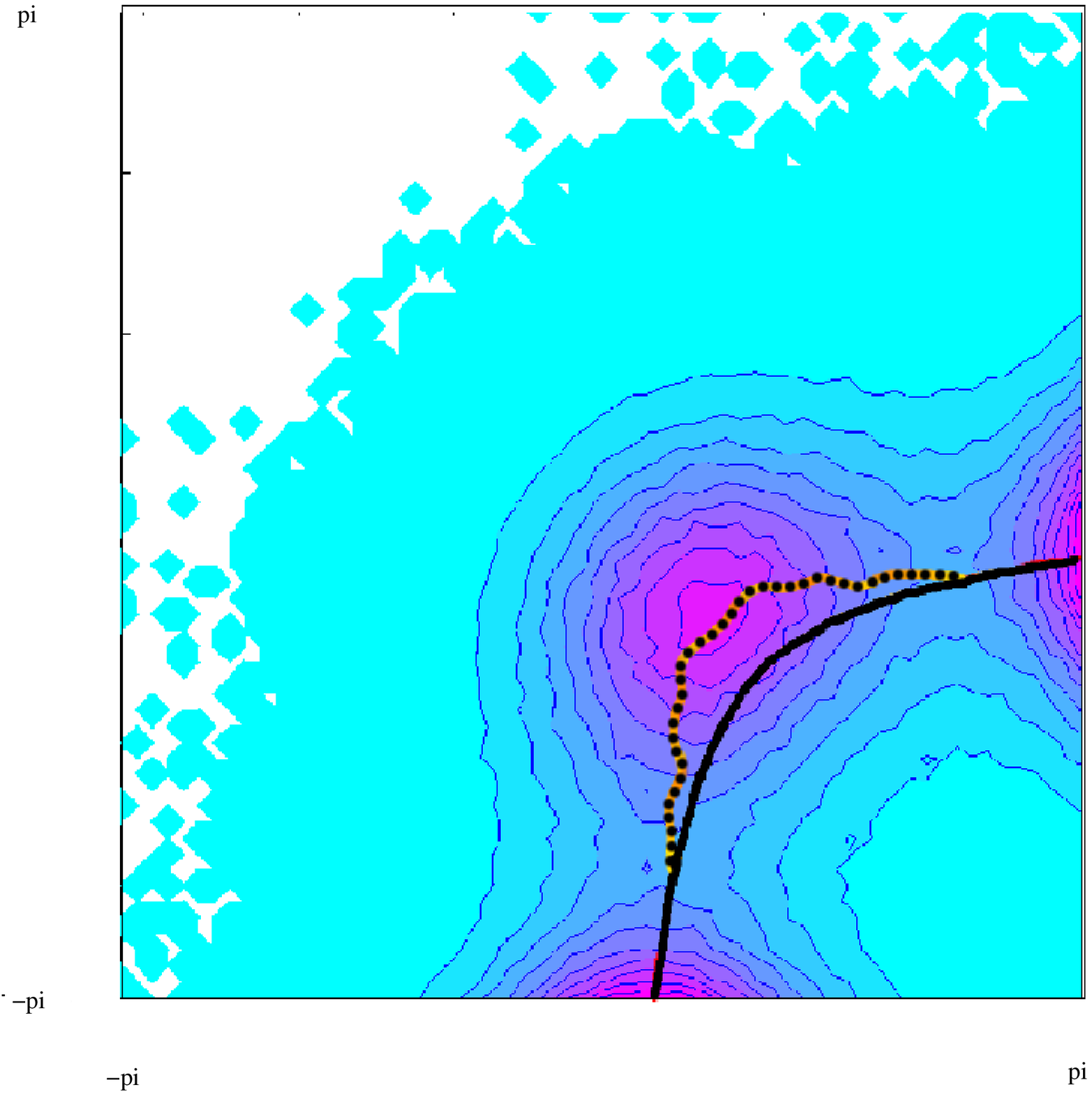}\\  
&$\qquad\;\theta_\Box^+$&&$\qquad\;\theta_\Box^+$
\end{tabular}   

\caption{Distribution of opposite plaquette values in cubes with one Dirac monopole at a) $\beta=0.5$ and b) $\beta=1.4$. The ridge of the distribution is indicated with a dotted line and compared to the curve from Fig.~\ref{fig:flux}b drawn in solid black. 
}\label{fig:plotvgl} 
\end{figure}

Due to the presence of Dirac strings, both diagrams reflect a $2\pi$ periodicity. To remove this dependence on the unphysical Dirac strings, we  use the reduced plaquette angle $\bar{\theta}_\Box\in[-\pi,\pi]$ in the further figures. The symmetry between monopoles and antimonopoles allows to restrict the analysis to monopoles only. Fig.\ref{fig:plotvgl} presents contour-plots for the distributions of ``$+1$-Dirac monopoles'' for $\beta=0.5$ to the left and $\beta=1.4$ to the right. By radial lines through $\bar\theta_\Box^\pm=(\pi,-\pi)$, which are in good approximation perpendicular to the lines of equal probability, we detect the ridge of the distribution (drawn in yellow). Its position is close to the curve of Fig.~\ref{fig:flux}b (drawn in black). 

The histograms in Fig.~\ref{fig:plotvgl} show maxima at $(\pi,-0.4)$, $(0.4,-\pi)$ and $(0.85,-0.85)$. Apart from statistical fluctuations the maximum line and the calculated curve of Fig.~\ref{fig:flux}b coincide except the aberration of the central maximum for monopoles in the center of a cube, which is shifted from $(\pi/3,-\pi/3)$ to $(0.85,-0.85)$. This shift is caused by the mutual influence of the three pairs of plaquettes in a cube due to the magnetic Gau{\ss} law, especially in the frequent situation when a monopole is located at the center of a plaquette, with a corresponding flux pair $(\pi,-0.4)$ or $(0.4,-\pi)$. Then the other two pairs share the remaining flux $(2\pi-\pi-0.4)/4\approx0.69$, and give data points at $(0.69,-0.69)$. Quantum fluctuations smear the peaks around $(\pi/3,-\pi/3)$ and $(0.69,-0.69)$, their superposition gives the observed maximum at $(0.85,-0.85)$.

\begin{figure}[h]   
\psfrag{b=0.5}[0][l][1][0]{$\beta=0.5$}
\psfrag{b=1.4}[0][l][1][0]{$\beta=1.4$}
\psfrag{left   -   plaquette   -   right}[c][c][1][0]{left$\quad\quad$-$\quad\quad$plaquette$\quad\quad$-$\quad\quad$right}   
\psfrag{probability density}[b][c][1][0]{``probability density''}   
\centering   
\includegraphics[scale=0.8]{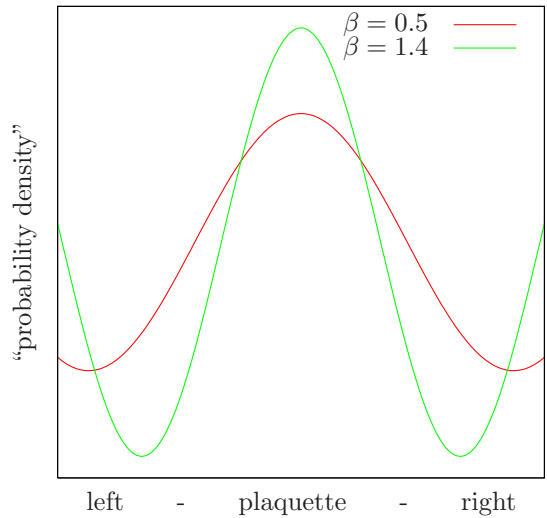}\\
\caption{``Probability density'' of plaquette pairs $\bar\theta_\Box^\pm$, which by Eq. (\ref{eq:flux}) are related to the position of monopoles within cubes, e.g. between left and right plaquette.}   \label{fig:resmpsxy} 
\end{figure}

Integrating the plaquette pair distribution along the above mentioned radial lines through $\bar\theta_\Box^\pm=(\pi,-\pi)$ of Fig.~\ref{fig:plotvgl} we get in Fig.~\ref{fig:resmpsxy} the probability for certain plaquette pairs as a function of the gradient angle of these radial lines. By Eq. (\ref{eq:flux}) these plaquette pairs are related to the position of monopoles within cubes as indicated in the title of the abscissa. With largest probability monopoles are located in the center of cubes. This is in accordance with earlier results~\cite{Bozkaya:2004tb}, where the monopole-antimonopole potential was found with local minima in the centers of cubes. Furthermore the probability is remarkably high for finding monopoles in the centers of plaquettes, where the potential energy is maximal and monopoles do not feel an accelerating force. 

Next we identify monopoles by the sinus flux definition and display histograms in the reduced plaquette angles $\bar{\theta}_\Box^\pm$. The flux measured by $\Phi_\mathrm{mag}=\sum_\Box\sin\bar\theta_\Box$ is maximal for monopoles in the center of a cube, $\Phi_\mathrm{mag}=6\sqrt{3}/2=5.2$. For monopoles located in the center of plaquettes $\Phi_\mathrm{mag}=2.9$. To get an overview of the dependence of the plaquette pair distribution on $\Phi_\mathrm{mag}$ we classify the monopoles according to the above numbers in four ranges of $\Phi_\mathrm{mag}$:

\begin{equation}    
\Phi_\mathrm{mag}=\sum_\Box\sin\bar{\theta}_\Box\in I=\{0,1.45\}, II=\{1.45,2.9\}, III=\{2.9,4\}\;\mathrm{or}\;IV=\{4,5.2\} \label{eq:sinranges}
\end{equation}

\begin{figure}[h]   
\centering   
\psfrag{1}{}
\psfrag{-1}{}
\psfrag{2}{}
\psfrag{-2}{}
\psfrag{0}[1][1][1][0]{$\hspace{-2mm}0$}
\psfrag{3}[1][1][1][0]{$\hspace{-2mm}\pi$}
\psfrag{-3}[1][1][1][0]{$\hspace{-5mm}-\pi$}
\psfrag{ 0}[1][b][1][0]{$\hspace{-1mm}0$}
\psfrag{ 3}[1][b][1][0]{$\hspace{-1mm}\pi$}
\psfrag{ -3}[1][b][1][0]{$\hspace{-3mm}-\pi$}
\begin{tabular}{cccc}     
0-Dirac monopoles:&&&\vspace{2mm}\\   
$\Phi_\mathrm{mag}\in I$:&$\Phi_\mathrm{mag}\in II$:&$\Phi_\mathrm{mag}\in III$:&$\Phi_\mathrm{mag}\in IV$:\vspace{6mm}\\
\includegraphics[scale=0.15]{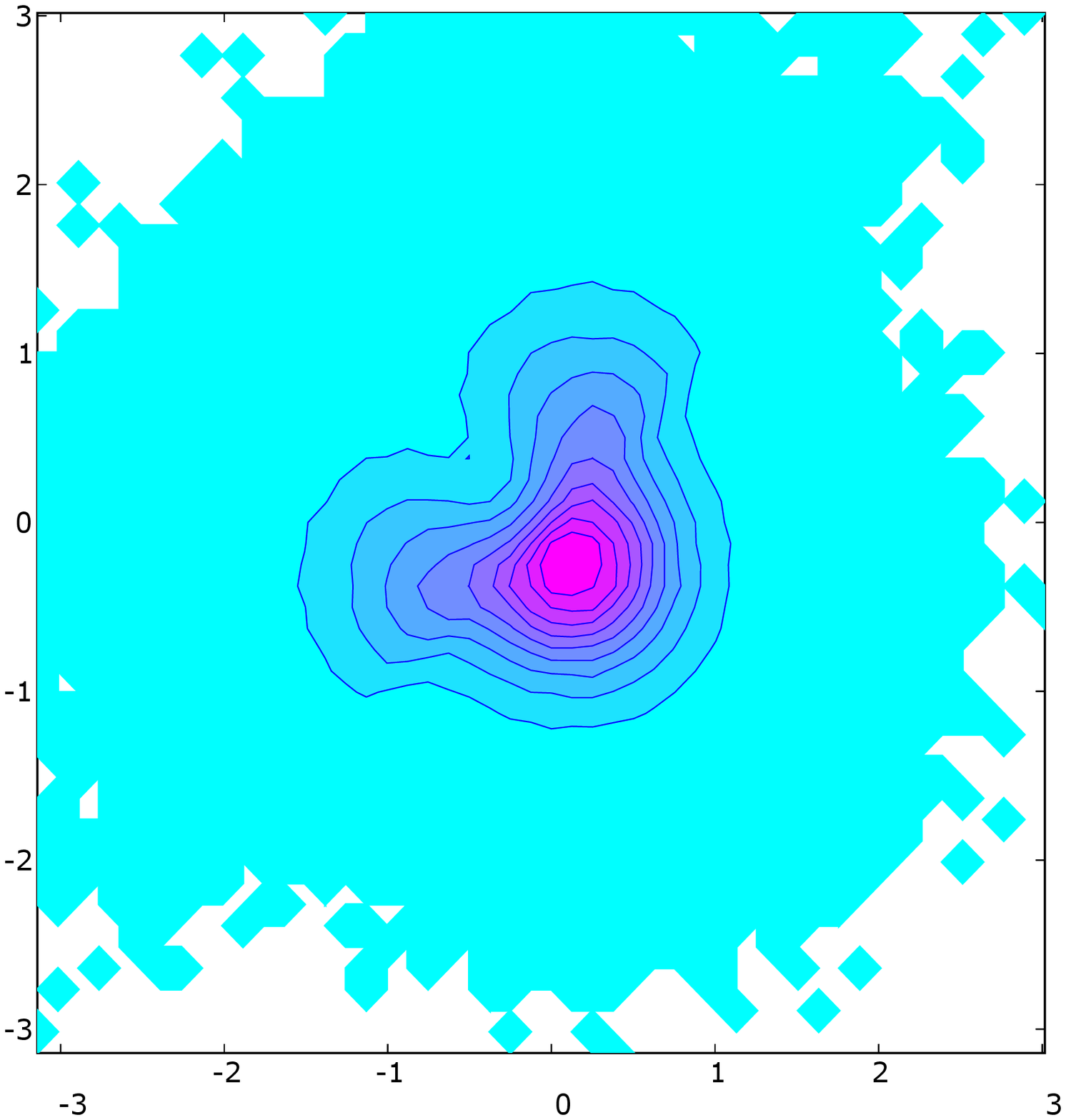}$\;$ & \includegraphics[scale=0.15]{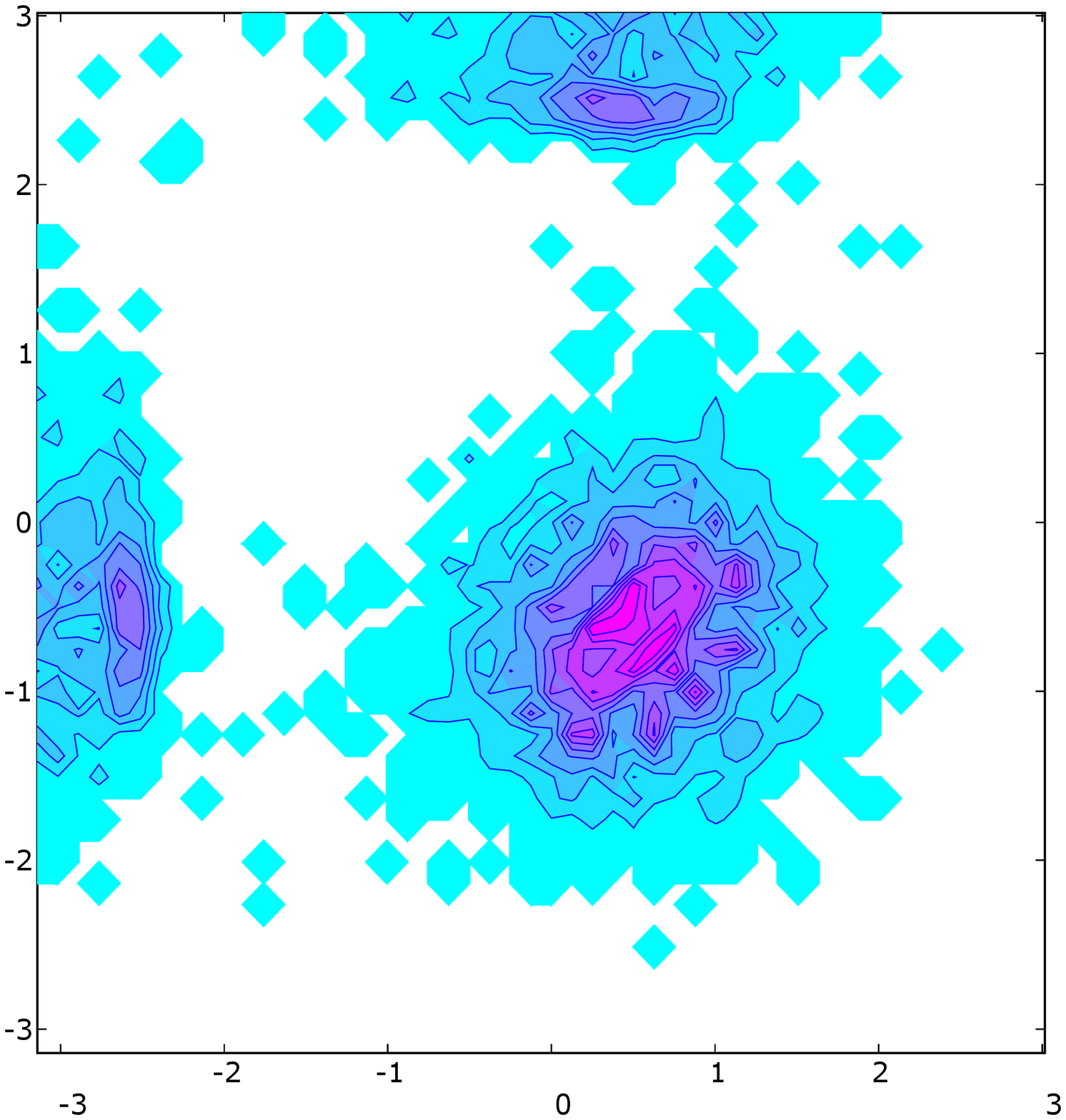}$\;$ & \includegraphics[scale=0.15]{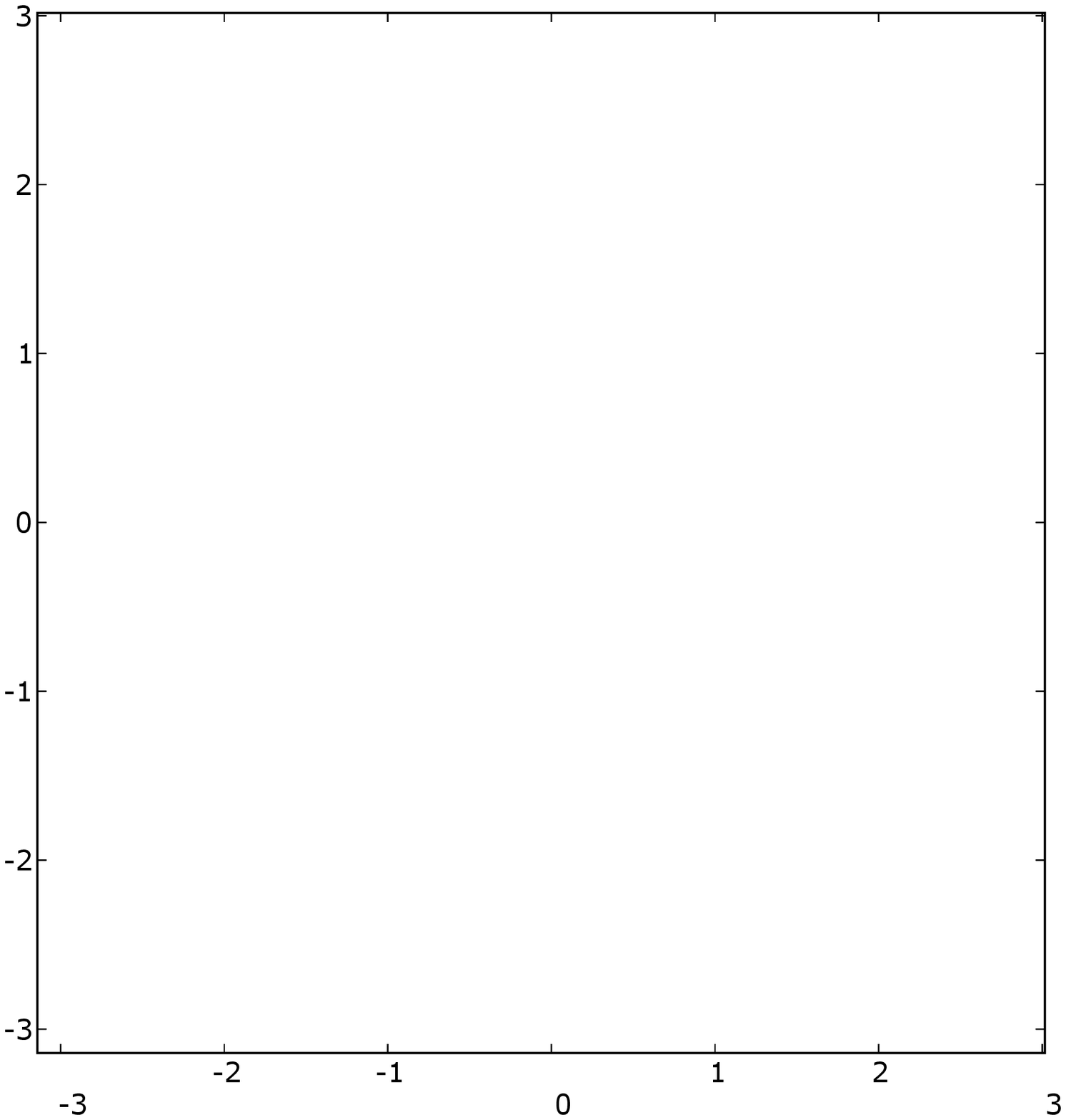}$\;$ & \includegraphics[scale=0.15]{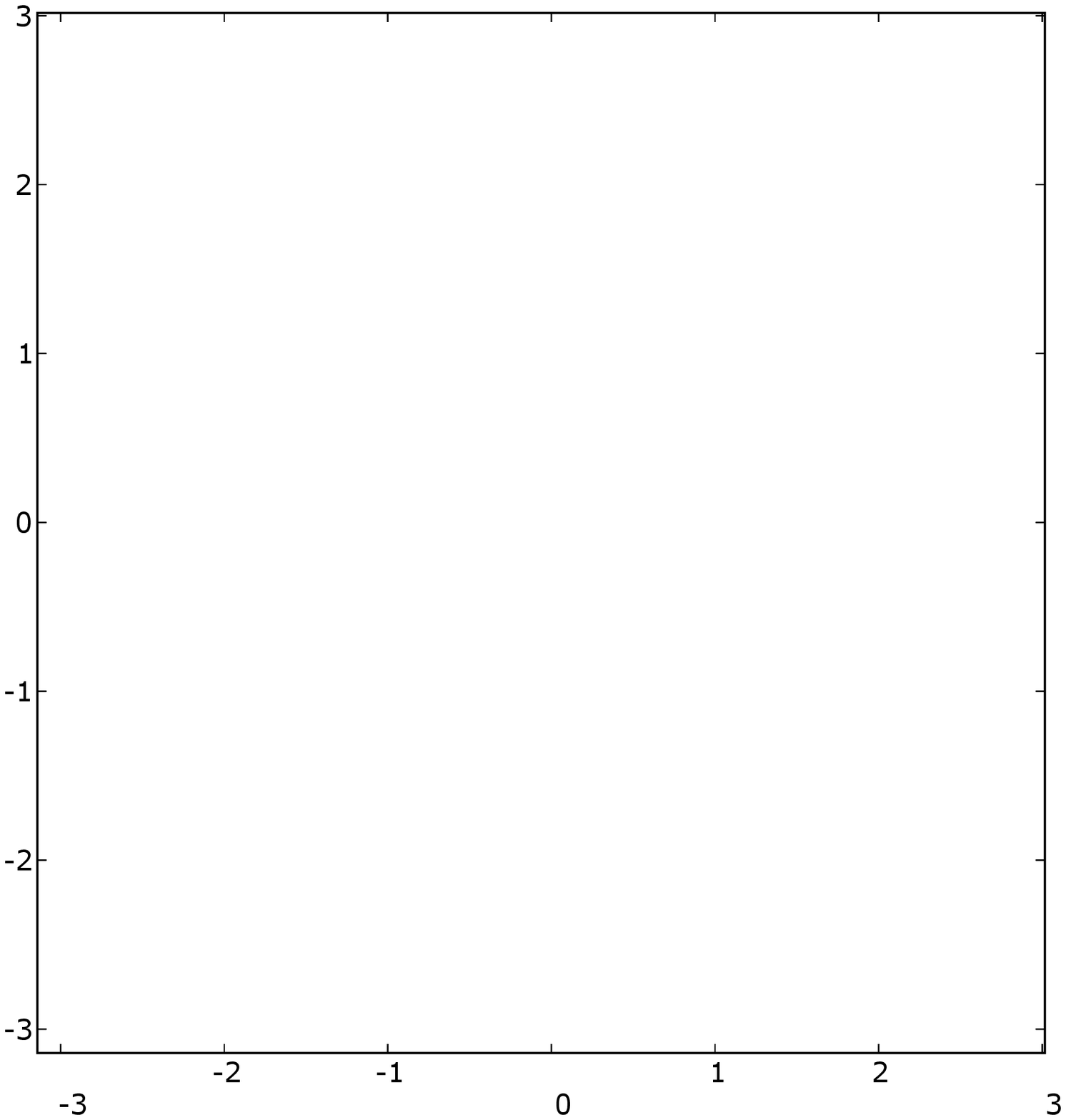}\vspace{3mm}\\     
tot.: 29.5\% & tot.: 0.2\% & tot.: 0\% & tot.: 0\% \vspace{3mm}\\   
1-Dirac monopoles:&&&\vspace{2mm}\\   
$\Phi_\mathrm{mag}\in I$:&$\Phi_\mathrm{mag}\in II$:&$\Phi_\mathrm{mag}\in III$:&$\Phi_\mathrm{mag}\in IV$:\vspace{6mm}\\
\includegraphics[scale=0.15]{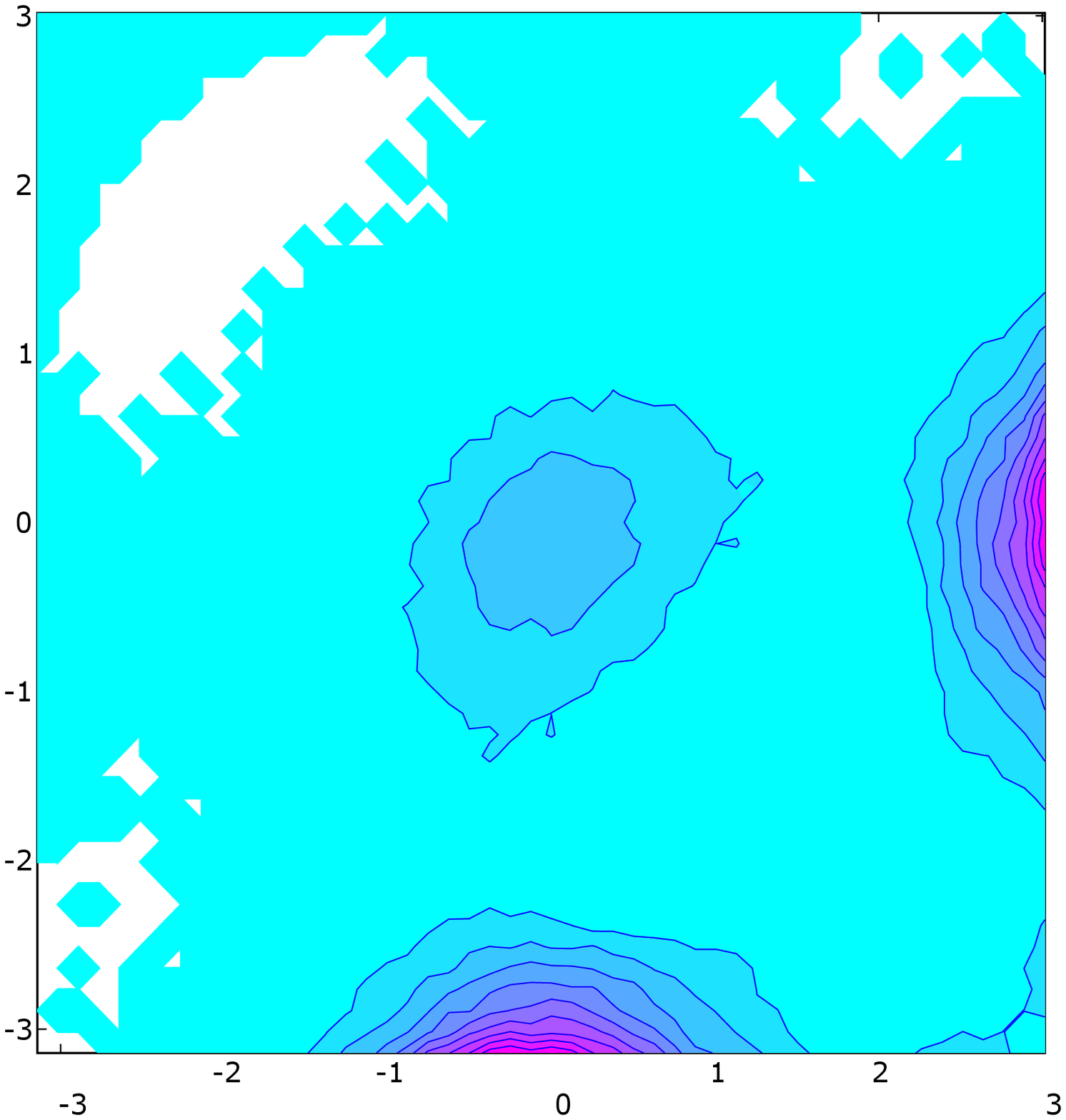}$\;$ & \includegraphics[scale=0.15]{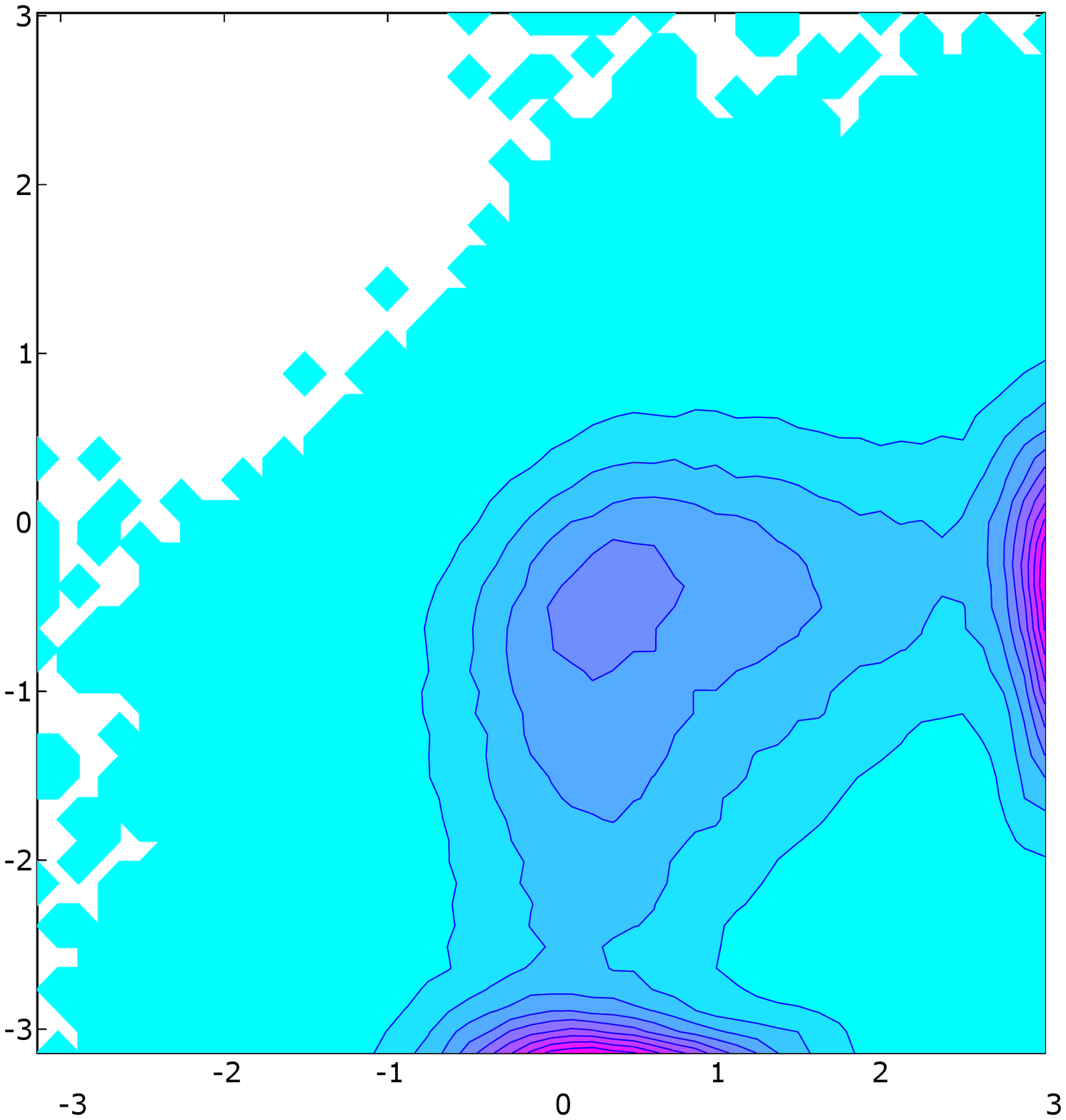}$\;$ & \includegraphics[scale=0.15]{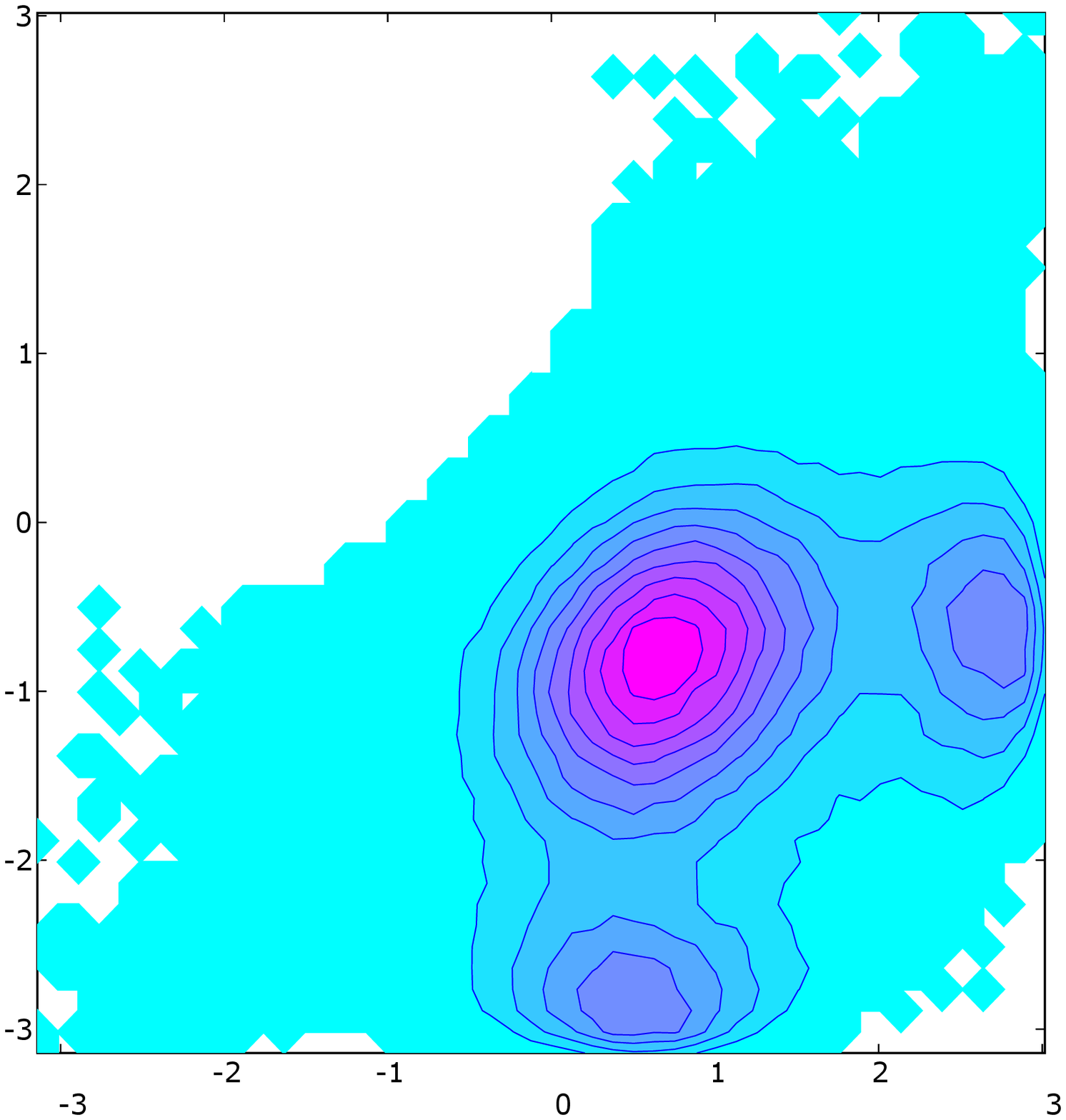}$\;$ & \includegraphics[scale=0.15]{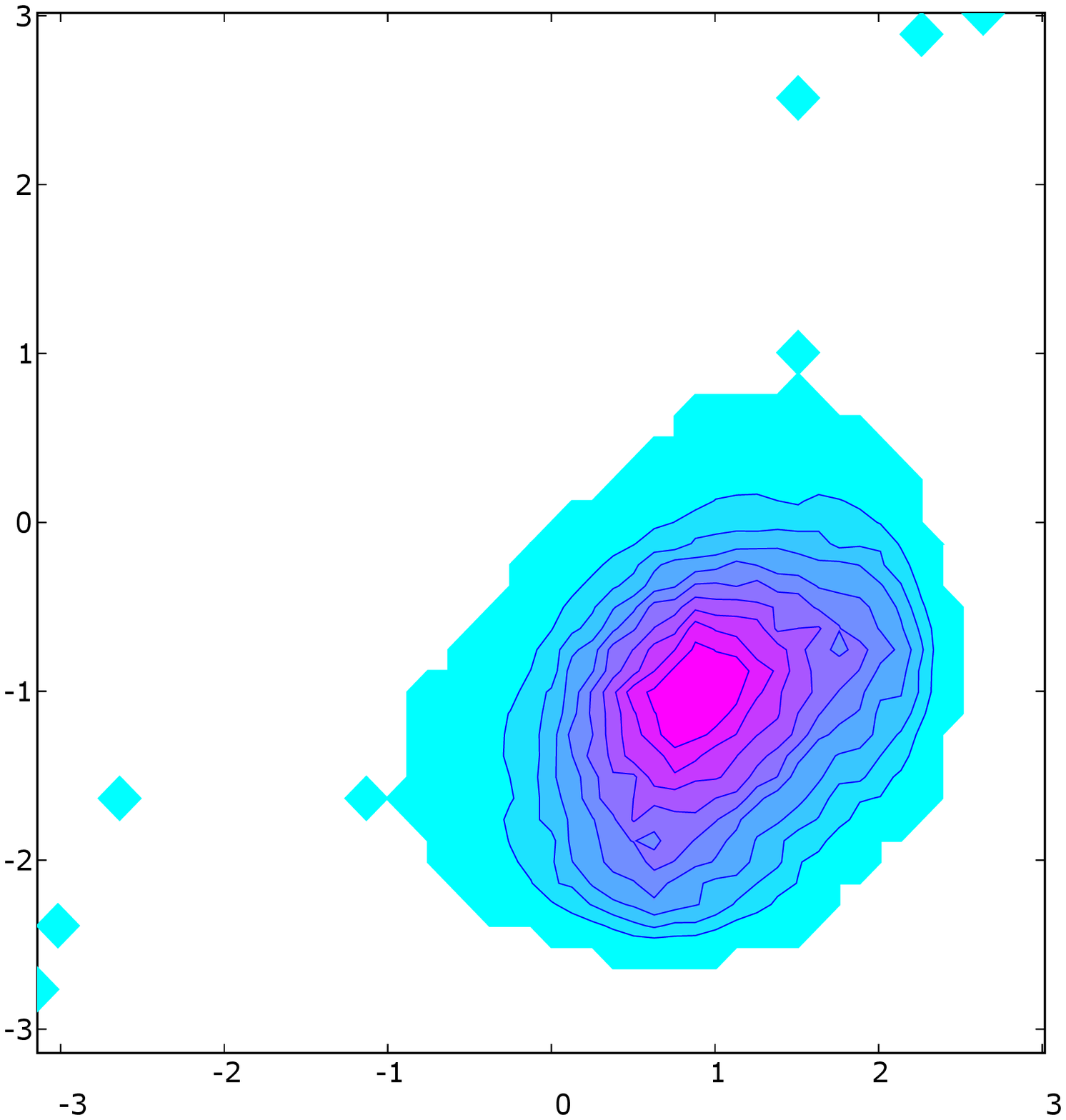}\vspace{3mm}\\     
tot.: 6.2\% & tot.: 36\% & tot.: 24.3\% & tot.: 3.8\% \\   
\end{tabular}   
\caption{Distributions of plaquette pairs $\bar{\theta}_\Box^\pm$ for various ranges of $\Phi_\mathrm{mag}=\sum_\Box\sin\bar{\theta}_\Box$ as defined in Eq.~(\ref{eq:sinranges}) for 0-Dirac monopoles (above) and 1-Dirac monopoles (below) at $\beta=1.4$. The relative number of cubes contributing to each plot is indicated by ``tot.:'', contour lines are relative to local maxima in 10\%-steps.}\label{fig:sinmps} 
\end{figure}

In Fig.~\ref{fig:sinmps} we present the distributions of plaquette pairs $\bar{\theta}_\Box^\pm$ for the ranges of $\Phi_\mathrm{mag}$ as defined in Eq.~(\ref{eq:sinranges}) for 0-Dirac monopoles (above) and 1-Dirac monopoles (below) at $\beta=1.4$. Every cube is therefore associated with one of the diagrams and contributes with three signals due to its three pairs of opposite plaquettes. 

For 0-Dirac monopoles we realize that most of the plaquette values of range $I$ are close to zero, only a few pairs in range $II$ (0.2\%) indicate the influence of a monopole which is closely outside of a cube. There are no cubes without Dirac monopoles with $\Phi_\mathrm{mag}>2.9$. For 1-Dirac monopoles and low $\Phi_\mathrm{mag}$ there are obviously a few cubes (6.2\%) where one of the plaquette values is close to $\pi$ and all other plaquettes are close to zero. This rather looks like a field fluctuation than a monopole. The monopoles are better developed in range $II$ (36\%) where they are located close to one of the plaquettes. In flux ranges $III$ and $IV$ most of the monopoles are located in the center of a cube. 

\begin{figure}
    \centering
    \psfrag{ 0}{$\scriptstyle0$}
    \psfrag{ 3.14159}{$\scriptstyle\pi$}
    \psfrag{-3.14159}{$\scriptstyle-\pi$}
    \psfrag{ 2.0944}{$\scriptstyle2\pi/3$}
    \psfrag{-2.0944}{$\scriptstyle-2\pi/3$}
    \psfrag{ 1.0472}{$\scriptstyle\pi/3$}
    \psfrag{-1.0472}{$\scriptstyle-\pi/3$}
	\includegraphics[width=0.5\textwidth]{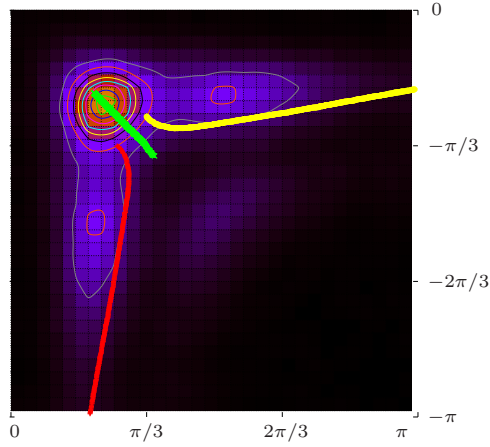}
	\caption{Distributions of plaquette pairs $\bar{\theta}_\Box^\pm$ for monopoles and antimonopoles of Monte-Carlo configurations after $200$ cooling steps. Further we show the plaquette pairs of Fig.~\ref{fig:coolhist} during the cooling procedure: in red (yellow) the plaquette pairs in dipole direction for monopoles (antimonopoles) move outwards and in green the orthogonal plaquette pairs move to the central maximum.}
	\label{fig:histcool}
\end{figure}

\begin{figure}
    \centering
	\includegraphics[width=0.8\textwidth]{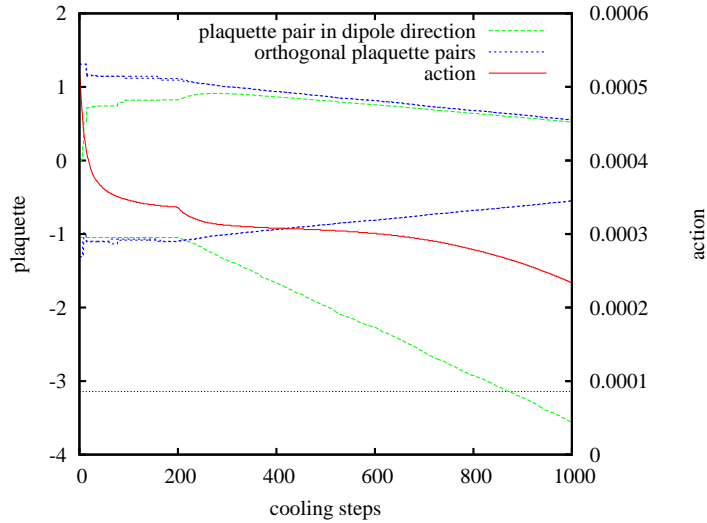}
	\caption{Plaquette pairs of a monopole-antimonopole pair and action during cooling. During the first 200 cooling steps two neighboring plaquettes were fixed to $-\pi/3$ resp. $\pi/3$, resulting in a monopole-antimonopole pair with distance 2. Releasing the constraints, the monopole-antimonopole pair shrinks and annihilates after 868 cooling steps, where the angle of the inner plaquette (green, dashed) crosses the horizontal line at $-\pi$.}
	\label{fig:coolhist}
\end{figure}

Finally, we look what happens during cooling in the confinement phase. We perform $200$ cooling steps with a spread of $0.4$ on $20$ Monte-Carlo configurations for $\beta=0.5$. The average monopole density is drastically reduced from $33.9\%$ to $2.3\%$ and we are left with a few monopole loops winding around the lattice. The distribution of plaquette pairs for monopoles and antimonopoles is plotted in Fig.~\ref{fig:histcool}. We see that there are no more monopoles located at plaquettes, they move away from the potential maxima. The central distribution maximum at $(\pi/3,-\pi/3)$ for uncooled configurations at $\beta=0.5$ moves towards $(0.7,0.7)$, which corresponds to the central maximum for uncooled configurations at $\beta=1.4$. There are indications that this behavior is due to the interaction of close monopole pairs, as we can see investigating pairs of static, magnetic monopoles at distance $d$. During cooling the monopole pairs at distances $d>2.5$ fall into a potential minimum at integer distances with plaquette pairs $(\pi/3,-\pi/3)$. If the initial distance $d<2.5$ no potential barrier keeps the monopole and antimonopole apart. Nevertheless at distance $d=2$ there is a metastable state (Fig. 5 in Ref.~\cite{Bozkaya:2004tb}) which slows down the shrinking process, see Fig.~\ref{fig:coolhist} where we plot the plaquette pairs of the monopole cubes during cooling. First we hold monopole-antimonopole pairs at distance $d=2.0$ by fixing the inner plaquette of both monopole cubes (lower green line) to $-\pi/3$ for 200 cooling steps with spread $0.4$. Then we continue cooling with a spread of $0.04$ and the action develops a plateau before the monopole-antimonopole pairs annihilate. The plaquette pairs for cooling steps larger than 200 are plotted within the distribution of Fig.~\ref{fig:histcool} until the monpole pairs annihilate after 868 cooling steps. The plaquette pairs in monopole direction (for monopoles in red and antimonopoles in yellow) move outwards indicating the annihilation process. The orthogonal plaquette pairs shift towards the maximum at (0.7,0.7). We conclude that cooling leads to approaching monopole loops with successive annihilation, during this process strong dipole fields are formed. 

\section{Conclusion}

We investigated the localization of magnetic monopoles detected by De Grand and Toussaint and the sinus flux definition methods. Magnetic monopoles defined by their charge distributions can take arbitrary positions on a discrete space-time lattice. However with increasing $\beta$ we find an increasing preference for certain positions within a cube. This monopole probability density correlates with the local extrema of the monopole-antimonopole potential found in Ref.~\cite{Bozkaya:2004tb}, minima in the centers of cubes and maxima in the centers of plaquettes, where monopoles do not feel an accelerating force either. During cooling the system of monopole loops seems to be influenced by attractive dipole forces.


\section*{Acknowledgment}

This study was partially supported by the Austria Science Fund (FWF) under grant P16910-N12.

\bibliographystyle{unsrt}
\bibliography{literatur}

\begin{thebibliography}{1}

\bibitem{Bozkaya:2004tb}
H.~Bozkaya, M.~Faber, P.~Koppensteiner, and M.~Pitschmann.
\newblock {Are there local minima in the magnetic monopole potential in compact
  QED?}
\newblock {\em Int. J. Mod. Phys.}, A19:5017--5026, 2004.

\bibitem{Dirac:1931kp}
Paul A.~M. Dirac.
\newblock Quantised singularities in the electromagnetic field.
\newblock {\em Proc. Roy. Soc. Lond.}, A133:60--72, 1931.

\bibitem{Wilson:1974sk}
Kenneth~G. Wilson.
\newblock Confinement of quarks.
\newblock {\em Phys. Rev.}, D10:2445--2459, 1974.

\bibitem{DeGrand:1980eq}
Thomas~A. DeGrand and Doug Toussaint.
\newblock {Topological excitations and Monte-Carlo simulation of Abelian gauge
  theory}.
\newblock {\em Phys. Rev.}, D22:2478, 1980.

\bibitem{Lang:1986ry}
C.~B. Lang and C.~Rebbi.
\newblock {Even and odd critical exponents of lattice QED}.
\newblock {\em Phys. Rev.}, D35:2510, 1987.

\bibitem{Skala:1996ar}
Peter Skala, Manfried Faber, and Martin Zach.
\newblock {Magnetic monopoles and the dual London equation in SU(3) lattice
  gauge theory}.
\newblock {\em Nucl. Phys.}, B494:293--311, 1997.

\bibitem{Zach:1995ni}
Martin Zach, Manfried Faber, Wolfgang Kainz, and Peter Skala.
\newblock {Monopole currents in U(1) lattice gauge theory: A Comparison to an
  effective model based on dual superconductivity}.
\newblock {\em Phys. Lett.}, B358:325--331, 1995.

\bibitem{Hoellwieser:2006}
{Roman H\"ollwieser}.
\newblock {Punktf\"ormige und ausgedehnte Monopole am Raum-Zeit-Gitter}.
\newblock Master's thesis, TU Wien, 2006.

\end{thebibliography}

\end{document}